\documentclass[aps]{revtex4}
\usepackage{graphicx}

\begin{document}
\title{Reexamination of optimal quantum state estimation of pure states}
\author{A.~Hayashi, T.~Hashimoto, and M.~Horibe}
\address{Department of Applied Physics\\
           Fukui University, Fukui 910-8507, Japan}

\begin{abstract}
A direct derivation is given for the optimal mean fidelity 
of quantum state estimation of a $d$-dimensional unknown pure state 
with its $N$ copies given as input, which was first obtained by 
M.~Hayashi in terms of an infinite set of covariant positive operator 
valued measures (POVM's) and 
by Bru{\ss} and Macchiavello establishing a connection to optimal 
quantum cloning. 
An explicit condition for POVM measurement operators
for optimal estimators is obtained, by which we construct optimal estimators 
with finite POVM using exact quadratures on a hypersphere.
These finite optimal estimators are not generally universal, 
where universality means the fidelity is independent of input states. 
However, any optimal estimator with finite POVM for $M(>N)$ copies is 
universal if it is used for $N$ copies as input.
\end{abstract}

\maketitle

\newcommand{\ket}[1]{|\,#1\,\rangle}
\newcommand{\bra}[1]{\langle\,#1\,|}
\newcommand{\braket}[2]{\langle\,#1\,|\,#2\,\rangle}
\newcommand{\bold}[1]{\mbox{\boldmath $#1$}}
\newcommand{\sbold}[1]{\mbox{\boldmath ${\scriptstyle #1}$}}
\newcommand{\tr}[1]{{\rm tr}\left[#1\right]}
\newcommand{\CH}{{\cal H}}
\newcommand{\CS}{{\cal S}}
\newcommand{\del}{\partial}

\section{Introduction}
One of the essential differences between quantum theory and classical 
theories, from the information theoretical point of view, is that  
an unknown quantum state cannot be copied exactly, which was formulated 
as the no-cloning theorem by Wootters and Zurek \cite{Wootters82}.

Suppose we are given $N$ identically prepared copies of an 
unknown state $\rho$ on a $d$-dimensional space $\CH_d$ and try to 
estimate the state $\rho$ as precisely as possible by some measurement. 
Since we cannot increase the number of copies by cloning the given unknown 
state, our performance surely depends on $N$, the number of copies given 
to us at the beginning. This is the problem of quantum state estimation, 
which has been studied since long ago \cite{Helstrom76,Holevo82}. 

Two important points in formulating quantum state estimation are a
prior distribution of the input states and a figure of merit to be 
optimized.
In this paper we assume that the state $\rho$ is pure and completely 
unknown in the sense that the state $\rho$ is distributed over all pure 
states in a unitary invariant way. As a figure of merit we take the 
fidelity defined as $\tr{\rho\rho'}=|\braket{\phi}{\phi'}|^2$, 
where $\rho=\ket{\phi}\bra{\phi}$ is the given input 
pure state and $\rho'=\ket{\phi'}\bra{\phi'}$ is the output pure state 
as a guess for $\rho$. 

The optimal mean fidelity in the case of qubit ($d=2$) pure state estimation 
was found to be $(N+1)/(N+2)$ by Massar and Popescu \cite{Massar95}. 
They also pointed out 
that the optimal value of the mean fidelity is achieved by a joint measurement
on the combined system of $N$ copies but not realized by repeated separate 
measurements on each copy. Conceptually this unexpected result should be taken 
seriously since the input is a simple uncorrelated $N$-fold tensor product 
$\rho^{\otimes N}$, though the improvement of joint measurement over separate 
measurement is relatively small (see also \cite{Bagan02}).
An algorithm for constructing an optimal and finite positive operator valued 
measure (POVM) has been given in 
\cite{Derka98}. Bagan {\it et al}. also discussed optimal and finite POVM's 
in two-dimensional case with a different approach \cite{Bagan01}.

Using the framework of covariant measurements \cite{Holevo82}, 
Masahito Hayashi studied the estimation problem in more general settings: 
in general dimensions $d$ and for a family of covariant error functions 
\cite{MHayashi98}. He showed that the error is minimized by the unique 
infinite covariant set of POVM's in both Bayesian and minimax approaches, 
provided that the error function is a monotone increasing function of 
$\tr{\rho\rho'}$. As for the mean fidelity he found the optimal value 
to be $(N+1)/(N+d)$. Bru{\ss} and Macchiavello also obtained the optimal 
mean fidelity by establishing a connection between 
optimal state estimation and optimal quantum cloning \cite{Bruss99}, 
the latter of which is another problem directly related to the no-cloning 
theorem. 
 
In optimal quantum cloning we are given $N$ identically 
prepared copies of quantum state $\rho$ on $\CH_d$ and try to 
produce a density matrix $R_\rho$ on $\CH_d^{\otimes M}$ in an 
approximation of $\rho^{\otimes M}$ as exactly as possible
\cite{Hillery97,Gisin97,Werner98,Keyl98}.
There are two kinds of figures of merit for approximate cloning. 
In the many-particle test the full fidelity $\tr{\rho^{\otimes M} R_\rho}$ 
is used for a figure of merit, 
whereas the one-particle reduced fidelity $\tr{\rho R_\rho}$ is 
employed in the single-particle test.
 
The general formula for the optimal many-particle fidelity as a function 
of $d$, $N$, and $M$ 
in the case of pure states was derived by Werner \cite{Werner98}.
It was also shown that the optimal fidelity is attained by the unique 
cloner. This unique optimal cloner was later shown to be also optimal with 
respect to the single-particle fidelity \cite{Keyl98}.

The connection established by Bru{\ss} and Macchiavello \cite{Bruss99} 
is the following (see also \cite{Gisin97}).
For given $N$ copies of a pure state, 
first employ the optimal cloner to produce infinite number of the 
best approximate copies, by which we can estimate the approximate copy 
as precisely as we want.
On the other hand applying the optimal estimator to the input first, 
we obtain the best approximate estimation of the input by which 
we can produce infinitely many copies of the same quality.
Thus they identified the optimal single particle 
fidelity in the large $M$ limit with the optimal mean fidelity of 
quantum state estimation.

For experimental implementation of POVM measurement, 
it is desirable that the number of outcomes of POVM measurement is finite. 
However, finite optimal POVM's for state estimation have been constructed 
only in the two dimensional case (qubit) 
\cite{Bagan02,Derka98,Bagan01,Latorre98}.
For general dimensions, the optimal fidelity was derived 
\cite{MHayashi98,Bruss99}, 
but finite optimal POVM's have not been discussed so far.
In this paper we will show that one can construct finite optimal POVM's 
in general dimensions. We also show that the finite optimal POVM 
may be chosen to be universal, where universality means the 
fidelity is independent of input states. 
These are the main results in this paper.

In Sec. II, we will first give a direct derivation of the optimal mean 
fidelity of quantum state estimation of a $d$-dimensional unknown pure state 
with its $N$ copies given as input.
Our main concern is whether the optimal fidelity can be achieved by 
a finite set of POVM's. 
Therefore we do not assume the covariance of 
measurement, since the covariance implies an infinite set of POVM's, 
when input states are specified by a set of continuous 
parameters as in the case considered in this paper.
We also avoid employing the optimal single-particle fidelity of 
approximate cloning, which is not straightforward to obtain. 
In Sec. III, we will study an explicit condition for POVM operators 
for optimal estimators. Using exact quadratures on a hypersphere, 
we establish the existence of a finite set of POVM's of optimal estimators.
Covariant measurements are universal whereas measurements 
with finite POVM are generally not, where universality means the 
fidelity is independent of the input states. 
In this respect we will also clarify 
the conditions under which measurements with finite POVM are universal and 
therefore also optimal in the minimax problem.

\section{Optimal mean fidelity}
Suppose we are given $N$ identically prepared copies of a randomly selected 
pure state $\rho=\ket{\phi}\bra{\phi}$ on a $d$-dimensional complex Hilbert 
space ${\cal H}_d$ and try to estimate the state $\rho$ as precisely as 
possible by some POVM measurement $\{E_a\}_{a=1}^A$ on $\rho^{\otimes N}$.
Since all inputs belong to the totally symmetric subspace of 
$\CH_d^{\otimes N}$, the completeness relation of the POVM can be written as 
$ \sum_{a=1}^A E_a = \CS_N$,
where $\CS_N$ is the projector onto this totally symmetric subspace.
With the outcome of the measurement labeled with "$a$", we infer that 
the state was a prespecified pure state $\rho_a = \ket{\phi_a}\bra{\phi_a}$.
Our task is to maximize the following mean fidelity:
\begin{eqnarray}
  F(N,d)= \sum_{a=1}^A 
     \left<
        \tr{E_a\rho^{\otimes N}} \tr{\rho_a\rho}
     \right>, \label{fidelity_def}
\end{eqnarray}
with respect to our strategy, the set of $\{E_a,\rho_a\}_{a=1}^A$. 
In the above equation $<\cdots>$ means an average over the input state 
$\rho$.

We assume the input state is distributed over all pure states on $\CH_d$ 
in a unitary invariant way. 
First let us fix an orthonormal basis $\{\ket{i},\ (i=1,\cdots,d)\}$ in 
${\cal H}_d$ and write a pure state $\ket{\phi}$ as 
$  \ket{\phi}=\sum_{i=1}^d c_i \ket{i} $,
where the coefficients $c_i$ satisfy the normalization condition 
$\sum_{i=1}^d c_i^* c_i = 1$.
We assume that $2d$-dimensional real vector $(\Re c_i,\Im c_i)_{i=1,\cdots,d}$
is uniformly distributed on $(2d-1)$-dimensional hypersphere. 
It is clear that the distribution defined above is independent of 
the reference basis. More precisely let 
$\{\ket{\widetilde{i}},\ (i=1,\cdots,d)\}$ be another orthonormal basis.
Then for any function $f$, the following can be easily shown:
\begin{eqnarray}
    \left< f\left(\sum_{i=1}^d c_i \ket{i}            \right) \right> =
    \left< f\left(\sum_{i=1}^d c_i \ket{\widetilde{i}}\right) \right>.
\end{eqnarray}

As shown in the appendix, the average of a product of the same number
of $c$'s and $c^*$'s is given by
\begin{eqnarray}
 \left< c_{i_1} c_{j_1}^*c_{i_2} c_{j_2}^* \cdots c_{i_l} c_{j_l}^* \right>
 = \frac{(d-1)!}{(d+l-1)!} 
      \left( \mbox{sum of all contractions between $i$'s and $j$'s} \right).
                             \label{contraction_formula}
\end{eqnarray}

Using this formula Eq.(\ref{contraction_formula}) and
writing a density operator for a pure state as 
$ \rho = \ket{\phi}\bra{\phi}=\sum_{i,j=1}^d c_i c_j^* \ket{i}\bra{j} $,
we obtain the following useful relation for the average of an $N$-fold 
tensor product of identical pure density matrices:
\begin{eqnarray}
  \left< \rho^{\otimes N} \right> = \frac{\CS_N}{d_N},
                    \label{rhoN_formula}
\end{eqnarray}
where the sum of all permutation operators divided by $N!$ is identified 
with ${\CS_N}$ and $d_N$ is  
the dimension of the totally symmetric subspace, which is given by 
$d_N=\tr{\CS_N}={}_{N+d-1}C_{d-1}$.

It should be noted that the relation Eq.(\ref{rhoN_formula}) is a consequence 
of the unitary invariance of the distribution of $\rho$, 
which can be seen in the following way. For any 
unitary $U$ on $\CH_d$ we have 
$   \left< U^{\otimes N} \rho^{\otimes N} U^{+\otimes N} \right>
  =\left< \rho^{\otimes N} \right>, $
implying that the operator $\left< \rho^{\otimes N} \right>$ 
on the totally symmetric subspace 
of $\CH_d^{\otimes N}$ commutes with $U^{\otimes N}$ for any $U$. 
Shur's lemma then requires that $\left< \rho^{\otimes N} \right>$ be 
proportional to ${\CS_N}$, 
since $U^{\otimes N}$ acts on the totally symmetric space irreducibly.
The proportional coefficient turns out to be $1/d_N$ by a 
trace argument. Thus we obtain the formula of Eq.(\ref{rhoN_formula}).

Going back to the mean fidelity Eq.(\ref{fidelity_def}), we first rewrite 
it as
\begin{eqnarray}
   F(N,d)= \sum_{a=1}^A 
     \left<
        \tr{E_a\rho^{\otimes (N+1)}\rho_a(N+1)}
     \right>, \label{fidelity_def2}
\end{eqnarray}
where the trace is taken over a total of $N+1$ subsystems and the operator 
$\rho_a(N+1)$ should be understood to act on the $(N+1)$th subsystem only; 
namely, for a single-particle operator $\Omega$ we use the following 
notation: 
$  \Omega(n) 
   \equiv 1^{\otimes (n-1)} \otimes \Omega \otimes 1^{\otimes (N+1-n)}$.
Using the formula Eq.(\ref{rhoN_formula}), we perform the integration over
$\rho$ to obtain
\begin{eqnarray}
   F(N,d)= \frac{1}{d_{N+1}}\sum_{a=1}^A 
             \tr{E_a\CS_{N+1}\rho_a(N+1)}. \label{fidelity_def3}
\end{eqnarray}
By tracing out the $(N+1)$th subsystem in the above equation, we finally 
obtain
\begin{eqnarray}
  F(N,d) &=& \frac{1}{(N+1)d_{N+1}} 
             \sum_{a=1}^A\tr{E_a
              \left( 1 + \sum_{n=1}^N \rho_a(n)  \right)
                        },   \nonumber \\
         & &  \label{fidelity_final} 
\end{eqnarray}
where we used the following relation which holds for any single-particle 
operator $\Omega$:
\begin{eqnarray}
  {\rm tr}_{N+1} \left[
          \CS_{N+1} \Omega(N+1)
                 \right]
 = \frac{1}{N+1} \CS_N \left(
            1 + \sum_{n=1}^N \Omega(n)
                       \right).
\end{eqnarray} 
 
Now it is easy to obtain an upper bound for the mean fidelity.
Since $ \rho_a(n) \le 1$, we have $\tr{E_a \rho_a(n)} \le \tr{E_a}$.
Applying this inequality to Eq.(\ref{fidelity_final}) and using the 
completeness of the POVM, we immediately find
\begin{eqnarray}
  F(N,d) \le \frac{d_N}{d_{N+1}}= \frac{N+1}{N+d}.
                            \label{upper_bound}
\end{eqnarray}

Equality in Eq.(\ref{upper_bound}) holds if and only if 
$\tr{E_a \rho_a(n)} = \tr{E_a}$ for $n=1,\cdots,N$, implying that 
$E_a$ is supported by the intersection of supports of $\rho_a(n)$, 
namely, $E_a$ is proportional to $\rho_a^{\otimes N}$.

Let us write $E_a$ as
\begin{eqnarray}
  E_a = d_N w_a \rho_a^{\otimes N}, \label{optimal_POVM}
\end{eqnarray}
where $w_a$ is a positive coefficient and the common factor $d_N$
is introduced for later convenience. The completeness of the POVM implies 
$   d_N\sum_{a=1}^A w_a \rho_a^{\otimes N} = \CS_N $.
Recalling the formula Eq.(\ref{rhoN_formula}), we conclude that the necessary 
and sufficient condition for the POVM that achieves the upper bound 
Eq.(\ref{upper_bound}) is given by
\begin{eqnarray}
  \sum_{a=1}^A w_a \rho_a^{\otimes N} = 
          \left< \rho^{\otimes N} \right>. \label{quadrature}
\end{eqnarray}

The right-hand side in this equation is the average of $\rho^{\otimes N}$ 
defined as a continuous integration over a hypersphere, 
whereas the left-hand side is the sum of a finite number of sample density 
operators $\rho_a^{\otimes N}$ with positive weights; 
namely, a continuous integration is replaced by a finite sum in 
Eq.(\ref{quadrature}), which is a standard technique of numerical 
integrations (quadrature). 
Though a quadrature is in general an approximation, 
it may be exact for a certain class of functions. 
For example, the quadrature with a trapezoidal rule is exact for any linear 
functions. 
In this sense the condition (\ref{quadrature}) means a quadrature 
on the hyper-sphere that is exact for $\rho^{\otimes N}$. 
Since quadratures with those properties exist as 
explicitly shown in the next section, we conclude that the optimal value 
of the mean fidelity is given by
\begin{eqnarray}
  F_{{\rm optimal}}(N,d) = \frac{N+1}{N+d}.  \label{optimal_fidelity}
\end{eqnarray}

\section{Finite set of POVM and universality}
In this section we first show that we can construct a finite set of POVM's 
that achieves the upper bound of the mean fidelity 
Eq.(\ref{upper_bound}) or equivalently there exists a
finite set $\{w_a,\rho_a\}_{a=1}^A$ which satisfies Eq.(\ref{quadrature}).
We write $\ket{\phi_a} = \sum_{i=1}^d c_i^a \ket{i}$ for 
$\rho_a=\ket{\phi_a}\bra{\phi_a}$. 
In terms of the expansion coefficient $c_i^a$, Eq.(\ref{quadrature}) 
is equivalent to
\begin{eqnarray}
  \sum_{a=1}^A w_a c_{i_1}^ac_{j_1}^{a*}c_{i_2}^ac_{j_2}^{a*}
                   \cdots c_{i_N}^ac_{j_N}^{a*}
 = \left<  c_{i_1}c_{j_1}^{*}c_{i_2}c_{j_2}^{*}
               \cdots c_{i_N}c_{j_N}^{*}
   \right>.
\end{eqnarray}
Equation (\ref{quadrature}) imposes no condition on products of different 
numbers of $c$'s and $c^*$'s.  To make the subsequent argument simpler, 
however, we assume that they are zero like their exact average value.
Then it suffices to show that there exists a quadrature with positive weights 
on the hypersphere $S^{2d-1}$    
that is exact for any polynomial of degree $2N$. 

For a point $\chi$ on $S^{m-1}$, $m\equiv 2d$, we 
write its polar coordinate parametrization as
\begin{eqnarray}
  \chi_1 &=& \cos\theta_1,    \nonumber\\
  \chi_2 &=& \sin\theta_1\cos\theta_2, \nonumber\\
  \chi_3 &=& \sin\theta_1\sin\theta_2\cos\theta_3,  \nonumber\\
  \vdots \nonumber\\
  \chi_{m-2} &=& \sin\theta_1\sin\theta_2\sin\theta_3\cdots\cos\theta_{m-2},
                                   \nonumber\\
  \chi_{m-1} &=& \sin\theta_1\sin\theta_2\sin\theta_3\cdots\sin\theta_{m-2}
                                                        \cos\phi,
                                   \nonumber\\
  \chi_{m} &=& \sin\theta_1\sin\theta_2\sin\theta_3\cdots\sin\theta_{m-2}
                                                        \sin\phi,
\end{eqnarray}
with the range of angle variables $0 \le \theta_i \le \pi$ and 
$0 \le \phi \le 2\pi$.
The integration measure is given by the standard form:
\begin{eqnarray}
  \int_0^\pi\!\!\! d\theta_1 \sin^{m-2}\theta_1
  \int_0^\pi\!\!\! d\theta_2 \sin^{m-3}\theta_2
  \cdots 
  \int_0^\pi\!\!\! d\theta_{m-2} \sin\theta_{m-2}
  \int_0^{2\pi}\!\!\! d\phi.
\end{eqnarray}

Let us consider the integral of 
$\chi_1^{\nu_1}\chi_2^{\nu_2}\cdots \chi_m^{\nu_m}$ with non-negative integers 
$\nu_\kappa$ which add up to $2N$. For each single integration we construct 
an $n$-point quadrature of the type 
$  \int\! dx f(x) = \sum_{\kappa=1}^n \omega_\kappa f(x_\kappa) $
with positive weights $\omega_\kappa$ 
that is exact for the functions under consideration.
We start with the $\phi$ integration:
\begin{eqnarray}
  \int_0^{2\pi}\!\!\! d\phi\, \cos^{\nu_{m-1}}\phi \sin^{\nu_m}\phi.
\end{eqnarray}
In the integrand we have the $(\nu_{m-1}+\nu_m)$th power of $e^{i\phi}$ 
or $e^{-i\phi}$ at most. Therefore a simple trapezoidal rule, 
$\omega_\kappa = 2\pi/n$ and $\phi_\kappa = 2\pi\kappa/n$, gives exact results 
provided that $\nu_{m-1}+\nu_m \le 2N < n$. We should remember that 
this integral vanishes unless both $\nu_{m-1}$ and $\nu_m$ are even.
 
Next we consider the $\theta_{m-2}$ integral
\begin{eqnarray}
  I_{m-2} \equiv
  \int_0^{\pi}\!\!\! d\theta\, \cos^{\nu_{m-2}}\theta 
                   \sin^{\nu_{m-1}+\nu_m+1}\theta,
\end{eqnarray}
where the subscript "$m-2$" of the variable $\theta$ is omitted. 
Note that we can assume 
$\nu_{m-1}+\nu_m$ is even since otherwise the whole integral is zero by 
the $\phi$ integration alone, which is exact, whatever wrong results other 
integrations produce. 
When $\nu_{m-1}+\nu_m$ is even, by setting $\cos\theta = x$ we obtain
\begin{eqnarray}
  I_{m-2} = 
  \int_{-1}^1\!\!\! dx\,x^{\nu_{m-2}} (1-x^2)^{(\nu_{m-1}+\nu_m)/2}.
\end{eqnarray}
Since the integrand is a polynomial of degree $\nu_{m-2}+\nu_{m-1}+\nu_m$,
we can use, for example, the Gauss-Legendre quadrature formula of weights 
$w_\kappa^{\rm g}$ and points $x_\kappa^{\rm g}$ (see \cite{Davis84} 
for example). In terms of variable $\theta$ the rule, 
$\omega_\kappa = w_\kappa^{\rm g}/\sin\theta_\kappa$ and 
$\theta_\kappa = \cos^{-1}x_\kappa^{\rm g}$, 
gives the exact result provided that 
$\nu_{m-2}+\nu_{m-1}+\nu_m \le 2N < 2n$.
Let us note that $I_{m-2} = 0$ unless $\nu_{m-2}$ is even.

We must examine one more integral, the $\theta_{m-3}$ integral:
\begin{eqnarray} 
  I_{m-3} \equiv
  \int_0^{\pi}\!\!\! d\theta\, \cos^{\nu_{m-3}}\theta 
                   \sin^{\nu_{m-2}+\nu_{m-1}+\nu_m+2}\theta.
\end{eqnarray}
By the same reason as in the case of $I_{m-2}$ we can assume that 
$\nu_{m-2}+\nu_{m-1}+\nu_m$ is even. 
Then the integration range can be enlarged 
to $[0,2\pi]$ so that an argument similar to that in the $\phi$ integration 
applies.
It turns out that the rule $\omega_\kappa=\pi/n$ and 
$\theta_\kappa=2\pi(2\kappa-1)/n$ is exact if 
$\nu_{m-3}+\nu_{m-2}+\nu_{m-1}+\nu_m+2 \le 2(N+1) < 2n$. If we changed 
the integration variable by $\cos\theta=x$, this rule would correspond to 
the Gauss-Tschebyscheff quadrature \cite{Davis84}.
The integral $I_{m-3}$ does not vanish only if $\nu_{m-3}$ is even.

It is clear that a similar argument also holds for remaining integrals; 
namely, for the integral $I_{m-i}$ the Gauss-Legendre (-Tschebyscheff) type 
of quadrature can be used when $i$ is even (odd). The weights $w_a$ of 
the whole integral are positive since they are given by the product of 
weights $\omega$'s of the single integrations. 
Thus we can conclude that there exists a
finite set $\{w_a,\rho_a\}_{a=1}^A$ which satisfies Eq.(\ref{quadrature}).  

Now that we have shown there exists a finite set of POVM's 
that achieves the optimal mean fidelity of Eq.(\ref{optimal_fidelity}), 
we study universality of finite optimal estimators. By universality we 
mean that the fidelity is independent of input states. 
The unaveraged fidelity of an optimal estimator for input $\rho^{\otimes N}$
can be written as
\begin{eqnarray}
  \sum_{a=1}^A \tr{E_a \rho^{\otimes N}}\tr{\rho_a\rho} 
    = d_N \sum_{a=1}^A w_a 
            \tr{\rho_a^{\otimes(N+1)}\rho^{\otimes(N+1)}}.
\end{eqnarray}
From this equation we find the fidelity is independent of $\rho$ if and 
only if 
$\sum_{a=1}^A w_a \rho_a^{\otimes(N+1)} = \CS_{N+1}/d_{N+1}$, 
namely
\begin{eqnarray}
  \sum_{a=1}^A w_a \rho_a^{\otimes(N+1)}= \left< \rho^{\otimes(N+1)} \right>.
                              \label{universal_quadrature}
\end{eqnarray}
This is a stronger condition than condition (\ref{quadrature}) that is 
required for optimal estimators for $N$ copies. 
Therefore optimal estimators are not generally universal, 
but any optimal estimators for $M(>N)$ copies are universal 
if it is used to estimate $N$ copies of an unknown state.

A closely related question to this issue is the following. 
Suppose that for given $N$ copies of an unknown pure state we first produce 
$M(>N)$ copies by the optimal cloner and then estimate the resulting 
state by the optimal estimator for $M$ copies. 
What is the fidelity of this apparently detourlike two-step estimation 
procedure?
Using the unique optimal cloner from $N$ to $M$ copies given by 
\cite{Werner98},
\begin{eqnarray}
   T(\rho^{\otimes N}) = \frac{d_N}{d_M} \CS_M
      \left( \rho^{\otimes N}\otimes 1^{\otimes (M-N)} \right) \CS_M,
\end{eqnarray}
and an optimal estimator $\{E_a,\rho_a\}_{a=1}^A$ for $M$ copies, we find 
that the unaveraged fidelity of the two-step estimation is optimal and 
universal:
\begin{eqnarray}
  \sum_{a=1}^A \tr{ E_a T(\rho^{\otimes N}) } \tr{\rho_a\rho}
 = d_N \sum_{a=1}^A w_a \tr{\rho_a^{\otimes N}\rho^{\otimes N}}
                        \tr{\rho_a\rho}
 = F_{{\rm optimal}}(N,d).
\end{eqnarray}

\section{Concluding remarks}
In this paper we gave a direct derivation of the optimal mean fidelity 
of pure state estimation in a way we find simpler than the original ones 
\cite{MHayashi98,Bruss99}. In order to show the existence of the 
optimal estimator with a finite POVM, we avoided the assumption of
covariance of the measurement and the use of connection to the optimal 
cloning fidelity.  
As a figure of merit we used the mean fidelity. 
It should be noted that the optimal fidelity is not changed 
if we take the infimum of fidelity for a figure of merit. 
As shown in the preceding section, in this minimax approach the condition for 
finite optimal estimators for $N$ copies is Eq.(\ref{universal_quadrature}), 
which gives the universal fidelity, instead of Eq.(\ref{quadrature}). 

We showed how to construct a finite set of POVM's for optimal estimators 
by the use of an exact quadrature on a hypersphere. 
We expressed the integration on hypersphere as a multiple of single 
integrations, for each of which we constructed an exact quadrature rule. 
But this procedure does not generally give the minimal set of points on 
the hypersphere, or equivalently the minimal set of POVM's. 
In the case of qubit the minimal set of POVM's has been studied for 
several values of $N$ \cite{Latorre98}. The weight $w_a$ obtained by our 
procedure depends on "$a$". It may be desirable to have a constant weight 
since $w_a$ is equal to the probability of finding outcome "$a$" for the 
random input considered in this paper; 
$w_a = < \tr{E_a \rho^{\otimes N}}>$. Exact quadratures with a constant 
weight for polynomials of degree $t$ on the hypersphere are called spherical 
$t$-designs. There is an existence result for all values of $t$ in any 
dimension \cite{Seymour84}, but explicit examples are in general not
straightforward to construct.

\begin{acknowledgments}
It is our pleasure to thank Masahito~Hayashi for his useful comments and 
for bringing the important paper \cite{MHayashi98} to our attention.
\end{acknowledgments}

\begin{appendix}
\section{}
In this appendix we sketch a derivation of the formula 
Eq.(\ref{contraction_formula}) based on a generating function 
for the readers convenience. 
In the text we considered the average of a function $f$ of a normalized 
complex vector $c = (c_1,c_2,\cdots,c_d)$ and its conjugate $c^+$:
\begin{eqnarray}
  < f(c,c^+) > = \frac{\int\! dc dc^+ f(c,c^+)}
                          {\int\! dc dc^+},
\end{eqnarray}
where
\begin{eqnarray}
     \int\! dcdc^+ = 
         \int_{-\infty}^{\infty} \prod_{i=1}^d d(\Re c_i) d(\Im c_i)
            \delta(c^+c-1).
\end{eqnarray}
It is convenient to introduce a generating function $G$ of 
$\lambda=(\lambda_1,\cdots,\lambda_d)$ and its conjugate $\lambda^+$:
\begin{eqnarray}
  G(\lambda,\lambda^+) = \int\! dcdc^+ e^{i(\lambda^+c + c^+\lambda)},
\end{eqnarray}
so that the average 
$\left< c_{i_1} c_{j_1}^*c_{i_2} c_{j_2}^* \cdots c_{i_l} c_{j_l}^* \right>$
is calculated as
\begin{eqnarray}
  \left< c_{i_1} c_{j_1}^*c_{i_2} c_{j_2}^*\cdots c_{i_l} c_{j_l}^* \right>
 = \left[ 
     \frac{\del}{\del \lambda_{i_1}^*}\frac{\del}{\del \lambda_{j_1}}
     \cdots
     \frac{\del}{\del \lambda_{i_l}^*}\frac{\del}{\del \lambda_{j_l}}
     G(\lambda,\lambda^+)
   \right]_ {\lambda=0}/ G(0).
\end{eqnarray}

First we express the $\delta$ function in the form of a Fourier transform as
$\delta(c^+c-1) = \frac{1}{2\pi}\int d\omega e^{i\omega(c^+c-1)}$.
Then the integration over $c$ and $c^+$ can be performed by a 
Gauss integral. The result is
\begin{eqnarray}
  G(\lambda,\lambda^+) = \frac{(i\pi)^d}{2\pi} \int\!\!d\omega\,
      \frac{1}{(\omega+i\epsilon)^d}
       e^{-i\frac{\lambda^+\lambda}{(\omega+i\epsilon)}} 
       e^{-i(\omega+i\epsilon)},
\end{eqnarray}
where $\epsilon$ is a small positive constant, which should go to zero 
in the end.
Expanding $e^{-i\frac{\lambda^+\lambda}{(\omega+i\epsilon)}}$ and 
performing the $\omega$ integration by a complex contour integral, 
we obtain
\begin{eqnarray}
  G(\lambda,\lambda^+) = \pi^d\sum_{n=0}^\infty \frac{1}{n!(n+d-1)!} 
                 (-\lambda^+\lambda)^n.
\end{eqnarray}
Now the formula of Eq.(\ref{contraction_formula}) can be shown as follows:
\begin{eqnarray}
 \left< c_{i_1} c_{j_1}^*c_{i_2} c_{j_2}^*\cdots c_{i_l}c_{j_l}^* \right>
 &=& \frac{(d-1)!}{l!(l+d-1)!}
     \frac{\del}{\del \lambda_{i_1}^*}\frac{\del}{\del \lambda_{j_1}}
     \cdots
     \frac{\del}{\del \lambda_{i_l}^*}\frac{\del}{\del \lambda_{j_l}}
     (\lambda^+\lambda)^l
                         \nonumber \\
 &=& \frac{(d-1)!}{(l+d-1)!} \left(
        \mbox{sum of all contractions between $i$'s and $j$'s} \right).
\end{eqnarray}
It is also easy to see that the average of any product of different number of
$c$'s and $c^*$'s vanishes.

\end{appendix}

\end{document}